\documentclass[reprint,showpacs,amsmath,amssymb,aps,pra,superscriptaddress,nobibnotes]{revtex4-1}
\usepackage[caption=false]{subfig}
\usepackage{graphicx}
\usepackage{dcolumn}
\usepackage{bm}
\usepackage{enumerate}
\usepackage{float}
\usepackage[ruled]{algorithm2e}   
\usepackage{tabularx}
\usepackage{threeparttable}
\usepackage{pifont}
\usepackage{placeins}
\usepackage[colorlinks,
linkcolor=red,
anchorcolor=blue,
citecolor=green
]{hyperref}
\begin{document}

\title{Deep Reinforcement Learning for Quantum Gate Control}

\author{Zheng An}
\affiliation{Institute of Physics, Beijing National Laboratory for
	Condensed Matter Physics,\\Chinese Academy of Sciences, Beijing
	100190, China}

\affiliation{School of Physical Sciences, University of Chinese
	Academy of Sciences, Beijing 100049, China}
\author{D. L. Zhou} \email[]{zhoudl72@iphy.ac.cn}

\affiliation{Institute of Physics, Beijing National Laboratory for
  Condensed Matter Physics,\\Chinese Academy of Sciences, Beijing
  100190, China}

\affiliation{School of Physical Sciences, University of Chinese
  Academy of Sciences, Beijing 100049, China}
\affiliation{Collaborative Innovation Center of Quantum Matter, Beijing 100190, China}
\affiliation{Songshan Lake Materials Laboratory, Dongguan, Guangdong 523808, China}
\date{\today}

\begin{abstract}
  How to implement multi-qubit gates efficiently with high precision
  is essential for realizing universal fault tolerant computing. For a
  physical system with some external controllable parameters, it is a
  great challenge to control the time dependence of these parameters
  to achieve a target multi-qubit gate efficiently and precisely. Here
  we construct a dueling double deep Q-learning neural network (DDDQN)
  to find out the optimized time dependence of controllable parameters
  to implement two typical quantum gates: a single-qubit Hadamard gate
  and a two-qubit CNOT gate. Compared with traditional optimal control
  methods, this deep reinforcement learning method can realize
  efficient and precise gate control without requiring any gradient
  information during the learning process. This work attempts to pave
  the way to investigate more quantum control problems with deep
  reinforcement learning techniques.
\end{abstract}

\pacs{03.67.Ac, 03.67.Lx, 07.05.Mh}

\maketitle

\section{Introduction \label{sec:intro}}

High fidelity quantum gate plays an essential role in achieving
quantum supremacy~\cite{supremacy} and fault-tolerant quantum
computing~\cite{PRESKILL1998}. In present days, the study of quantum
control has developed a series of methods in practice, such as nuclear
magnetic resonance experiments~\cite{Vandersypen}, trapped
ions~\cite{Islam2011,Jurcevic2014}, superconducting
qubits~\cite{Barends2016}, and nitrogen vacancy
centers~\cite{Zhou2016}. Further, based on gradient or evolutionary
algorithms, the development of control algorithms provides robust
control strategies and have been intensively used. However, it is hard
to get such high-quality gates under limited control resources with a
precise choice of the control signal, like time-discretization of the
fields or fixed amplitude. In a previous work~\cite{Larocca2018},
under certain limitations, the quantum control landscape was
non-convex but will get dumped in the vicinity of quantum speed limit
time. Even though the result of the topology of quantum control
landscapes has been intensively tested and
studied~\cite{Nielsen,PhysRevA.86.013405,Nanduri}, it is hard to
minimize errors of some quantum systems. In addition, these problems
can be generalized to hard quantum control
problems~\cite{Zahedinejad}. All these limitations are hard to be
solved with common quantum-control techniques but meaningful for being
discussed in the physical world.

On the other hand, machine learning, already explored as a tool in
many aspects of physics~\cite{Hezaveh2017,Biamonte2017}, provides a
complete paradigm to achieve analysis of various quantum systems
~\cite{Biamonte2017,Carleo2017,Carrasquilla2017,Nieuwenburg2017}. With
tremendous aspects studied in ML, reinforcement learning (RL) has been
a focus on the study of artificial intelligence agent to interact with
the real world. Equipped with deep neural network, the deep RL
techniques has revolutionized traditional optimal control which
provides efficient, precise, and robust performance. Further
empowered by advanced optimization techniques, the artificial
intelligence agent is able to solve high-dimensional optimization
problems such as video games and go~\cite{atari,alphago,alphazero}.
Recently, researchers have begun to utilize some RL algorithms in the
quantum control studies~\cite{Bukov2017,Niu2018}. The novel RL
algorithm provides advanced optimization techniques which are able to
solve more difficult optimization problems.

In this article, we investigate the traditional quantum gate control
problem where an efficient strategy for preparing high fidelity
quantum gate proposed by an artificial intelligence agent. With deep
RL, we propose a framework to connect optimal decision making of the
underlying quantum dynamics with state-of-the-art RL techniques. In
particular, within the present framework, the agent performs optimal
discrete, sequential controls to get two typical quantum gates: a
single-qubit Hadamard gate and a two-qubit CNOT gate. The results
provide a general way to investigating the quantum control problem
with deep RL techniques.

The rest of this paper is structured as follows. In
Sec.\ref{sec:bang}, we briefly overview our quantum gate control
model. In Sec.\ref{sec:phase}, we present some relative RL algorithms
and the DDDQN method for two quantum gate control models. In Sec.
\ref{sec:result} and \ref{sec:conclued}, we show the numerical results
and draw our conclusions.

\section{Bang-bang control model to implement quantum gates \label{sec:bang}}

In this section, we give a bang-bang control model to implement
quantum gates, which explains the physical problems we solve in this
paper.

We consider a quantum system whose Hamiltonian is
\begin{equation}
  H(\vec{\epsilon}(t))={H}_{d} + {H}_{c}(\vec{\epsilon}(t)),
\end{equation}
where the term ${H}_d$, called the drifted Hamiltonian, is the free
evolution part of the Hamiltonian ${H}(\vec{\epsilon}(t))$. Another part of
the Hamiltonian, ${H}_c(\vec{\epsilon}(t))$, called the control Hamiltonian,
is under control by some time dependent external parameter vector
$\vec{\epsilon}(t)$.

In our bang-bang control protocol, our total control time $T$ is
fixed, which is divided into $N$ short time periods with the same
duration $\delta t = T/N$. In the $i$-th time period with
$(i-1)\delta t\le t\le i \delta t$ ($1\le i\le N$), the control parameter vector is
constant, i.e. $\vec{\epsilon}(t)=\vec{\epsilon}_{i}$, where the control parameter
vector $\vec{\epsilon}_{i}$ are selected from a set
$\mathcal{A}(\vec{\epsilon})$ of $d$ possible choices. The unitary evolution operator
in the $i$-th time period is
\begin{equation}
  \label{eq:2}
  U(i\delta t, (i-1)\delta t; \vec{\epsilon}_{i}) = e^{-i H(\vec{\epsilon}_{i}) \delta t}.
\end{equation}
When all the $N$ control parameter vectors
$\{\vec{\epsilon}_{1},\vec{\epsilon}_{2},\ldots,\vec{\epsilon}_{N}\}$ are selected, the unitary
operator at time $T$ is determined by the iterative equations
\begin{align}
  U(i \delta t) & = U(i\delta t, (i-1)\delta t;\vec{\epsilon}_{i}) U((i-1)\delta t), \\
  U(0) & = I,
\end{align}
where $I$ is the identity operator in the Hilbert space of our system.

Our aim is to select the parameter vectors
$\{\vec{\epsilon}_{1},\vec{\epsilon}_{2},\ldots,\vec{\epsilon}_{N}\}$ to make the unitary
operator $U(T)$ approximate the target unitary gate $U_{f}$ as well as
possible, which is formulated by maximizing the fidelity
\begin{equation}
  \label{eq:3}
  \mathcal{F}(T) = \max_{\vec{\epsilon}_{1},\vec{\epsilon}_{2},\ldots,\vec{\epsilon}_{N}} \mathcal{F}(T;\vec{\epsilon}_{1},\vec{\epsilon}_{2},\ldots,\vec{\epsilon}_{N})
\end{equation}
with the fidelity
\begin{equation}
  \mathcal{F}(T;\vec{\epsilon}_{1},\vec{\epsilon}_{2},\ldots,\vec{\epsilon}_{N}) = \left|
    \frac{\mathrm{Tr}\{{U}_{f}^{\dagger} {U}(T)\}}{D} \right|^{2},
\end{equation}
where $D$ is the dimension of the Hilbert space. We observe that
$\mathcal{F}(T;\vec{\epsilon}_{1},\vec{\epsilon}_{2},\ldots,\vec{\epsilon}_{N})\in[0,1]$, and that
$\mathcal{F}(T;\vec{\epsilon}_{1},\vec{\epsilon}_{2},\ldots,\vec{\epsilon}_{N})=1$ if and only if
${U}(T)$ is equal to ${U}_{f}$ up to a phase factor.

In particular, the size of the set of the parameter vectors is
$d^{N}$, which implies that it is impossible to exhaustively searching
the optimal parameter vector sequence for a large $N$.

Here we focus on two typical target quantum gates, one is the Hadmard
gate, the other is the CNOT gate.

\subsection{Hadamard gate}

When the target quantum gate is the single
qubit Hadmard gate
\begin{equation}
  U_{f}=\frac{1}{\sqrt{2}}\begin{pmatrix}
    1&1\\1&-1
  \end{pmatrix},
\end{equation}
we consider a two-level system whose Hamiltonian is 
\begin{equation}
  H(\epsilon(t))=\sigma_{z}+\epsilon(t)\sigma_{x}
  \label{ham},
\end{equation}
where $\sigma_{z}$ and $\sigma_{x}$ are Pauli matrices, and
$\epsilon(t)$ is a real control parameter. This simple model has been widely
applied in quantum physics, e.g., it describes the non-adiabatic
transition~\cite{1932}, the Landau-Zener-Stuckelberg
interferometry~\cite{Shevchenko2010} and the Kibble-Zurek
mechanism~\cite{Zurek}.

Based on the Pontryagin maximum principle, we take the set of $d=2$
possible control parameter $\mathcal{A}(\epsilon)\in \{\pm4\}$ in our bang-bang protocol.

\subsection{CNOT gate}

When the target quantum gate is the CNOT
gate
\begin{equation}
  U_{f}=\begin{pmatrix}
    1&0&0&0\\0&1&0&0\\0&0&0&1\\0&0&1&0
  \end{pmatrix},
\end{equation}
we consider the Hamiltonian
\begin{equation}
  \begin{split}
    H(\epsilon(t))= & \sigma_{z}^{(1)} \otimes \sigma_{z}^{(2)} +
    \epsilon_{1}(t)\sigma_{x}^{(1)}\otimes\mathbb{I}^{(2)} + \epsilon_{2}(t)\mathbb{I}^{(1)}\otimes\sigma_{x}^{(2)}\\
    & + \epsilon_{3}(t)\sigma_{y}^{(1)}\otimes\mathbb{I}^{(2)} +
    \epsilon_{4}(t)\mathbb{I}^{(1)}\otimes\sigma_{y}^{(2)},
  \end{split}
  \label{ham_c}
\end{equation}
where $\mathbb{I}$ is the $2\times2$ identity matrix, and
$\vec{\epsilon}(t)=(\epsilon_{1}(t),\dots,\epsilon_{4}(t))$ is a $4$ component parameter
vector.

Similarly as in the case of the Hadmard gate, we take the set of
$d=16$ possible choices of the parameter vector as
\begin{equation}
  \label{eq:4}
  \mathcal{A}(\vec{\epsilon}) = \{(\epsilon_{1},\epsilon_{2},\epsilon_{3},\epsilon_{4}) \text{ with } \epsilon_{i}\in\{\pm 4\}\}. 
\end{equation}

\section{Deep reinforcement learning methods~\label{sec:phase}}

In this section, we show how to apply the deep RL
to approximately solve the maximization problem specified by
Eq.~(\ref{eq:3}) in our bang-bang control quantum gate implementation
protocol. To this end, we firstly review the necessary concepts in
deep RL methods, especially the framework of the
dueling double deep Q-learning neural network, which is adopted in our
problem. Then we show how to combine our bang-bang control protocol
with the deep RL methods.

\subsection{Reinforcement learning \label{RL}}

RL is a kind of ML method in which an intelligent agent aims to find a
series of actions on a given environment to optimize its performance
by delayed scalar rewards received~\cite{sutton}.

The problem of RL is described as a finite Markov decision
process~\cite{sutton}. At time $t=0$, the state of the environment is
$S_{0}$, and the agent chooses an action $A_{0}$. At time $t=1$, the
state of the environment becomes $S_{1}$ after the action $A_{0}$, and
the environment also gives a scalar reward $R_{1}$. Then the agent
chooses an action $A_{1}$, and repeats the above procedure.  
In general, this Markov process is described as a
state-action-reward sequence
\begin{equation*}
  S_{0}, A_{0}, R_{1}, S_{1}, A_{1}, R_{2} , \dots
\end{equation*}
For a finite Markov decision process, the sets of the
states, the actions and the rewards are finite. The total discounted
return at time $t$
\begin{equation}
  G_{t}=\sum_{k=0}^{\infty}\gamma^{k}R_{t+k+1},
\end{equation}
where $\gamma$ is the discount rate and $0\le\gamma\le1$. 

In RL, the agent selects the actions according to a policy $\pi$, which
is specified by a conditional probability of selecting an action $A$
for each state $S$, denoted as $\pi(A|S)$. The task of the agent is to
learn an optimal policy $\pi_{\ast}$, which maximizes the expected
discounted return
\begin{equation}
  V_{\pi}(s) = E_{\pi}(G_{t}|S_{t}=s),
\end{equation}
where $E_{\pi}$ denotes the average expectation under the policy
$\pi$.

It has been shown that the optimal policy $\pi^{*}$ exists and can be
found iteratively as follows. Let us introduce the value of
state-action function, the conditional discount return
\begin{equation}
  \label{eq:1}
  Q_{\pi}(s,a) = E_{\pi}(G_{t}|S_{t}=s,A_{t}=a).
\end{equation}
If we have a policy $\pi$, then we calculate the value of state-action
function $Q_{\pi}(S,A)$. For each state $s$, we take an action
maximizing the value of state action $Q_{\pi}(s,A)$, which forms our new
policy $\pi^{\prime}$. Then we calculate the value of state-action function
$Q_{\pi^{\prime}}(S,A)$. Repeating the above procedure until the new policy
equals the updated one, which is the optimal policy $\pi^{*}$ we are
looking for.

Another well known method to get the optimal policy $\pi^{\ast}$ is
the Q-learning~\cite{Watkins1989},  an off-policy temporal-difference
control algorithm defined as
\begin{equation}
  Q(S_{t},A_{t})  \leftarrow Q(S_{t},A_{t}) + \Delta Q
    \label{Q}
\end{equation}
with
\begin{equation}
  \label{eq:5}
 \Delta Q = \alpha[R_{t+1}+\gamma \max_{a}Q(S_{t+1},a)-Q(S_{t},A_{t})],
\end{equation}
where $\alpha$ is the step size parameter.

\subsection{Dueling Double Deep Q-learning}

\begin{figure*}
  \centering
  \begin{minipage}[b]{0.45\textwidth}
    \includegraphics[width=0.98\columnwidth{},keepaspectratio]{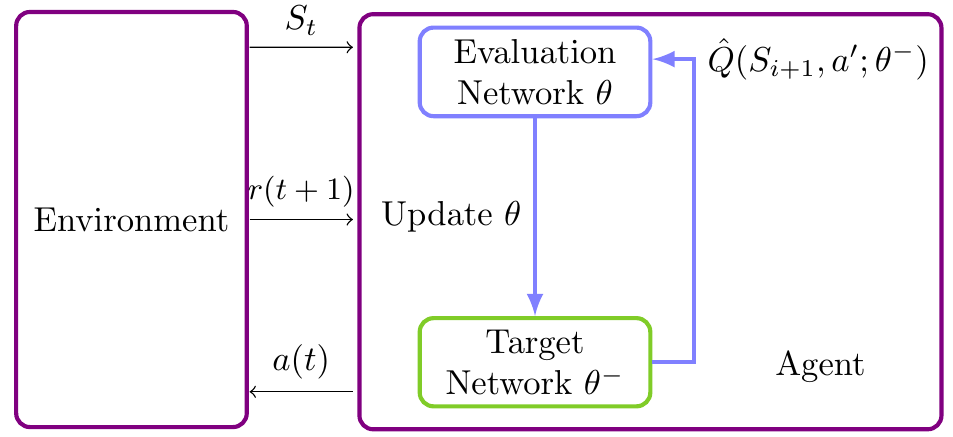}
    \caption{An overview of the deep RL: at each
      time step of training, the evaluation network of the agent
      proposes a control action of $a(t)$, the environment takes the
      proposed action and evaluates gate of Eq. (\ref{eq:2}) for
      time duration $\delta t$ to obtain a new unitary gate $U_{t+1}$ and
      calculates the reward of Eq. (\ref{reward}) , both of which are
      fed into the RL agent. The evaluation network of the agent is
      updated with the loss function of Eq. (\ref{loss}) by
      backpropagation. With fixed numbers of steps, the agent updates
      the parameters of the target network by transferring the
      parameters of evaluation network.}\label{reinforce}
  \end{minipage}\qquad
  \begin{minipage}[b]{0.45\textwidth}
    \includegraphics[width=1\columnwidth{}]{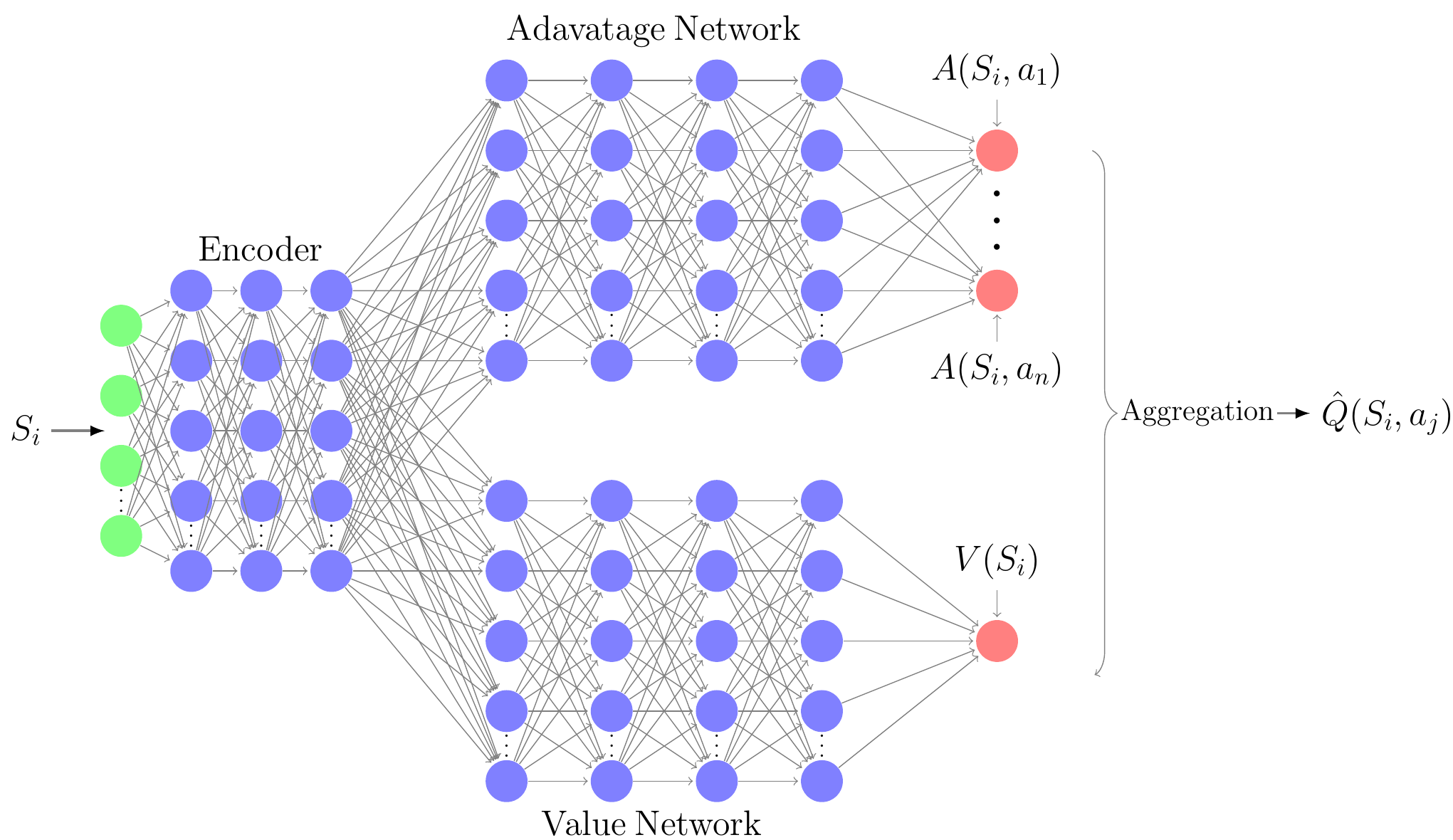}
    \caption{The deep neural network architecture of our agent: For
      each state fed into the neural network, the encoder extracts the
      information of the state for further calculation. The value
      network and advantage network get the information of the encoder
      to calculate the value of the state for each action. Based on
      Eq. (\ref{adv}), the neural network aggregates the value of the
      state and the advantage of action by the state to get the
      state-action values $\hat{Q}(S_{i},a_{j})$.}
    \label{net}
  \end{minipage}
\end{figure*}

	In this section, we introduce the Dueling Double Deep Q-learning
Neural Network (DDDQN), which will be used in our quantum gate control
problem. The advantage of this method has been discussed in previous research~\cite{dueling}.

First, we begin by introducing the double Q-learning
method~\cite{double} in the training of our agent. As shown in
Fig.~\ref{reinforce}, the agent consists of the evaluation network and
the target network with the same architecture. The evaluation network
evaluates the state-action value ${Q}(S,A; \theta)$, and the target network
evaluates the TD target ${Q}(S,A; \theta^{-})$. At each learning step, we
fed the agent with a minibatch of experiences
$\{S_{t},A_{t},R_{t+1},S_{t+1}\}$ with the prioritized experience
replay (PER) method~\cite{PER}. The state $S_{t}$ is fed into the
evaluation network to calculate the state-action value
${Q}(S_{t},A_{t};\theta)$. At the same time, the target network is to
calculate $\max\limits_{a'}{Q}(S_{t+1},a';\theta^{-})$ in Eq.~(\ref{loss}).
At the end of each step of training, the evaluation network is updated
through the back-propagation by minimizing the loss. Based on
Eq.~(\ref{Q}), the loss is the mean square error (MSE) of the
difference between the evaluation ${Q}(S_{t},A_{t};\theta)$ and the target
$\max\limits_{a'}{Q}(S_{t+1},a';\theta^{-})$
\begin{equation}
  \rm{loss}=\rm{MSE}((R_{t+1} + \gamma\max\limits_{a'}{{Q}}(S_{t+1},a';\theta^{-}))-{Q}(S_{t},A_{t};\theta)).
  \label{loss}	
\end{equation} 
During the learning episodes (see Fig.~\ref{reinforce}), the agent
updates the parameters of the target network
$\theta^{-} \rightarrow \theta$ to make better decisions.

Further, the detailed architecture of each network in our agent is
shown in Fig.~\ref{net}. Each network is consisted of three
parts: an encoder, an advantage network and a value network. The
encoder extracts information about the states $S_{t}$ for the next two
neural networks. Based on the Q-learning, the state-action value
${Q}(S_{t},A_{t})$ represents the expected return for the agent to
select the action $A_{t}$ on the state $S_{t}$ of the environment. In
the architecture of the dueling network~\cite{dueling} in deep RL, we
decompose the state-action value as
\begin{equation}
  Q(S_{t},A_{t})= A(S_{t},A_{t})+V(S_{t}),
  \label{adv}
\end{equation}
where $V(S_{t})$ is the state value for each state, and
$A(S_{t},A_{t})$ is the advantage for each action. The state value
$V(S_{t})$ is calculated by the advantage network, and the advantage
of action $A(S_{t},A_{t})$ is calculated by the value network. Then we
combine these two values to get an estimate of $Q(S_{t},A_{t})$
through an aggregation layer.

\subsection{Quantum gate control with DDDQN}
\label{sec:quantum-gate-control}

To apply the reinforcement ML to our bang-bang control protocol, we
need to build a map between their concepts. The state of the
environment at time $t$ is
\begin{equation}
  \begin{split}
    S_{t}= U(t \delta t) = \{\Re(U_{ij}(t \delta t)), \Im(U_{ij}(t \delta t))\},
  \end{split}
\end{equation}
where $U_{ij}(t \delta t)$ is the matrix element of $U(t\delta t)$, and
$\Re, \Im$ mean taking the real part and the imaginary part. The action
the agent at time $t$ can take
\begin{equation}
  \label{eq:6}
  a(\vec{\epsilon}) = U(t \delta t, (t-1)\delta t; \vec{\epsilon}).
\end{equation}
Note that the action does not depend on time $t$. The reward of the
agent received in each step is
\begin{equation}
  R_{t}=\begin{cases}
    0, & t\in\{0,1,\dots,N-1\}\\
    - \mathcal{L}(\mathcal{F}(T;\vec{\epsilon}_{1},\vec{\epsilon}_{1},\ldots,\vec{\epsilon}_{N})), & t=N
  \end{cases}
  \label{reward}
\end{equation}
where $\mathcal{L}(\mathcal{F})$ is the logarithmic infidelity,
$\mathcal{L}(\mathcal{F})=\log_{10}(1-\mathcal{F})$. In other words, the
agent will not get a reward immediately, but at time $N$.

Our algorithm for quantum gate control with DDDQN is given in
Algorithm~\ref{DRL}.

\begin{algorithm}
  \caption{\centering Deep RL for quantum gate control}
  \label{DRL}
  \SetNlSkip{0.25em} \SetInd{0.5em}{1em}
  Initialize memory R to empty\;
		
  Randomly initialize the evaluation network with random weights $\theta$\;
		
  Randomly initialize the target network with random weights $\theta^{-}$\;
		
  \For{episode= 0, M} {Initialize $s_{0}$ , $s_{0}=f(U_{0})$\;
			
    \For{$t=0, \dots, t_{N}$}{With probability $\epsilon$ select a random
      action $a_{t}$, otherwise $a_{t}=argmax_{a}{Q}(s_{t},a;\theta)$\;
			
      Execute action $a_{t}$ and observe the reward $r_{t+1}$, and the
      next state $s_{t+1}$\;
			
      Store experience $ e_{t}= (s_{t}; a_{t}; r_{t+1}; s_{t+1})$ in
      R\;
		
      \If{$t=t_{N}$}{ Sample minibatch of experiences $e_{i}$ with PER
        method\;
		
        Set $y_{i}=\left\{\begin{array}{lc}
                            r_{i+1}&\text{if $t_{i+1}=t_{N}$ }\\
                            r_{i+1}+\gamma
                            \mathop{\arg\max}_{a'}{Q}(s_{t+1},a';\theta^{-})&\text{otherwise}
                          \end{array}\right.
                        $\
				
                        Update $\theta$ by minimizing
                        $\text{loss}=(y_{i}-{Q}(s_{t},a_{i};\theta))^{2}$\;
                      }\ Every C times of learning, set
                      $\theta^{-}=\theta$\; }\ }\
\end{algorithm}

\section{Numerical results\label{sec:result}}
 
In this section, we give the numerical results of the logarithmic
infidelity $\mathcal{L}$ with target gates being the single-qubit
Hadamard gate and the two-qubit CNOT gate from the deep reinforcement
learning. To show the effectiveness of our deep RL method, we also
calculate the logarithmic infidelity with three different algorithms:
gradient ascent pulse engineering (GRAPE), differential evolution
(DE), and genetic algorithm (GA). We then present our analysis of the
performance of our deep RL algorithm against the other three
algorithms.
 
  \begin{figure}
    \includegraphics[width=0.98\columnwidth{},keepaspectratio]{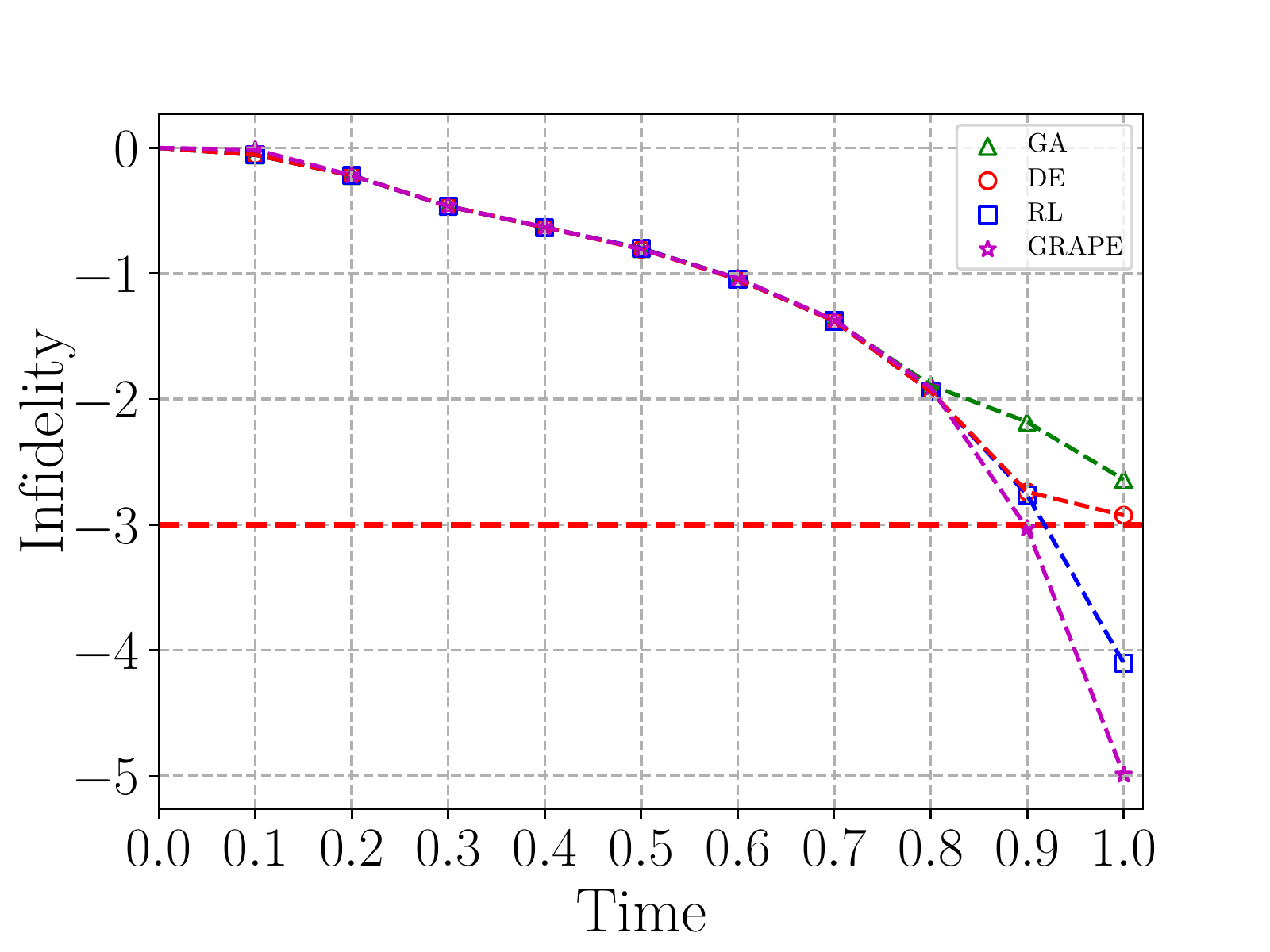}
    \caption{\label{fig:compfid}Best infidelities of preparing a
      single-qubit Hadamard gate in different evolution time T. The
      markers correspond to the algorithms RL (blue $\Box$), GRAPE(purple
      \ding{73}), DE(red $\circ$) and GA (green $\triangle$). The time step
      $N=28$ for different $T$. Here we set 400 iterations for GRAPE
      DE and GA, 100000 training episodes for RL.}
  \end{figure}
  \begin{figure}
    \includegraphics[width=0.98\columnwidth{}]{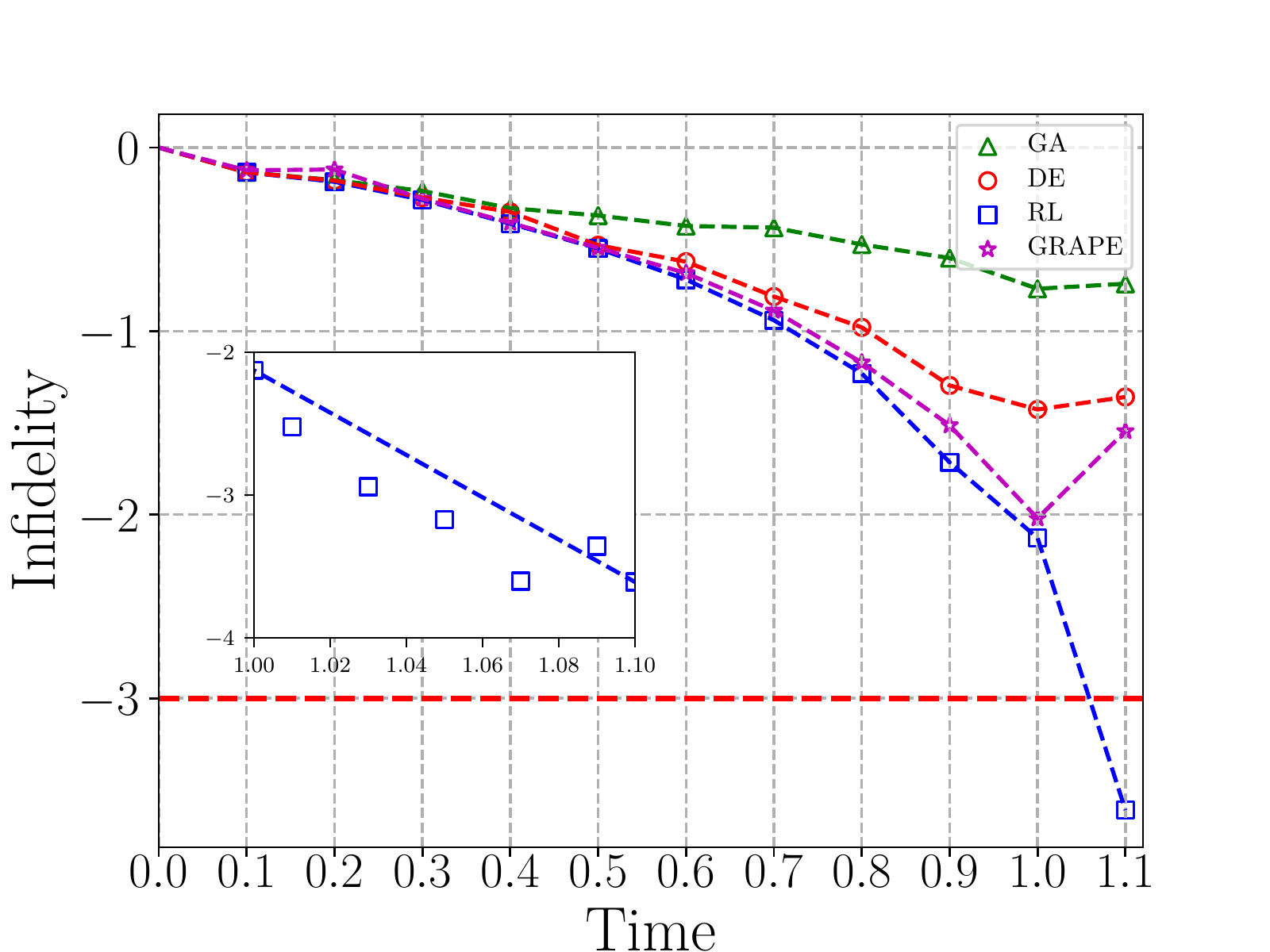}
    \caption{{\label{fig:compinf}}Best infidelities of preparing a
      CNOT gate in different evolution time T. The markers correspond
      to the algorithms RL (blue $\Box$), GRAPE(purple \ding{73}), DE(red
      $\circ$) and GA (green $\triangle$). The time step $N=38$ for different
      $T$. Here we set 5000 iterations for GRAPE DE and GA, 150000
      training episodes for RL.}
  \end{figure} 

  Fig.~\ref{fig:compfid} shows the minimal logarithmic infidelities of
  preparing a single-qubit Hadmard gate in different evolution time $T$
  with different algorithms. For $T<0.8$, the results on the
  infidelities from the four algorithms agree well. At $T=0.9$, the
  results on the logarithmic from RL and DE agree well, which is
  better than that from GA, and worse than that from GRAPE\@. At
  $T=1.0$,  the infidelity obtained from
  RL and GRAPE abruptly decrease, which possibly implies that the
  speed limit time of the problem is in the region $[0.9,1.0]$. In
  particular, these two algorithms find protocols to achieve
  infidelity $\mathcal{L}<-3$ (red line) or fidelity
  $\mathcal{F}>99.9\%$ at $T=1.0$. While GRAPE has the best
  performance out of the four methods, the algorithm requires the
  fidelity gradients at all time, and it is not readily accessible through
  experimental measurements. Further, GRAPE allows
  for the control field $\epsilon(t)$ to take any value in the interval
  $[-4,4]$.
  
  In Fig.~\ref{fig:compinf}, we compare the results of the CNOT gate
  control task from the four algorithms. Similar as in the previous
  task, all the algorithms perform well for $T<0.4$. RL, DE and GRAPE
  find optimal protocols in the time region $0.4<T<0.9$, but the
  performance of GA is poor for $T>0.4$. After $T=0.9$, only RL and
  GRAPE find optimal protocols, and the results of our RL agent are
  better than that of the GRAPE\@. Notice that at $T=1.1$, the landscape
  seems to get dumped for the problem and all the algorithms except RL
  get trapped. Like the state transfer problem
  ~\cite{Alexandrec,Bukov2017}, we believe this region may have a similar
  phase transition phenomenon and traditional algorithms are hard to
  maintain good performance. However, our RL agent ignores the dumped
  landscape and finds good protocols compared with other algorithms. To
  investigate the performance of RL in this region, we plot detailed
  results in the inset of Fig.~\ref{fig:compinf}. The results show
  that the agent has good and robust performance in the region.

  \section{CONCLUSION \label{sec:conclued}}
  
	In this article, we apply the deep RL to explore the fast and
high-precision quantum gate control problem. The quantum gate
control problem is then mapped into a deep RL algorithm.
Further, we build an RL agent to solve the quantum optimal
control problem. We compare the numerical results among the
four different algorithms on two typical quantum gate control
problems. Our results demonstrate that the artificial
intelligent is able to effectively learn the optimal control
schemes in approximating the target quantum gates.
The success of our agent lies in its suitability for
	solving discrete action problems and its state of art RL
	technique of balancing explore and exploit.

The numerical results show that the performance of
	deep RL is robust and efficient in implementing arbitrary
	single and two qubit gates. However, there are still some
	challenges to extend RL algorithms to multi-qubit control
	problem. The main challenge needs to solve is that the
	control space will grow exponentially with the increase of
	qubit number. We hope that our approach can inspire more
	applications of deep RL methods in the quantum control
	domain.
 
 \begin{acknowledgements}
   This work is supported by NSF of China (Grant Nos. 11475254 and
   11775300), NKBRSF of China (Grant No. 2014CB921202), the National
   Key Research and Development Program of China (2016YFA0300603).
 \end{acknowledgements}
 
 \appendix
 \section{Hyper-Parameters and Learning Curves}

\begin{figure}
 	\includegraphics[width=0.99\columnwidth{},keepaspectratio]{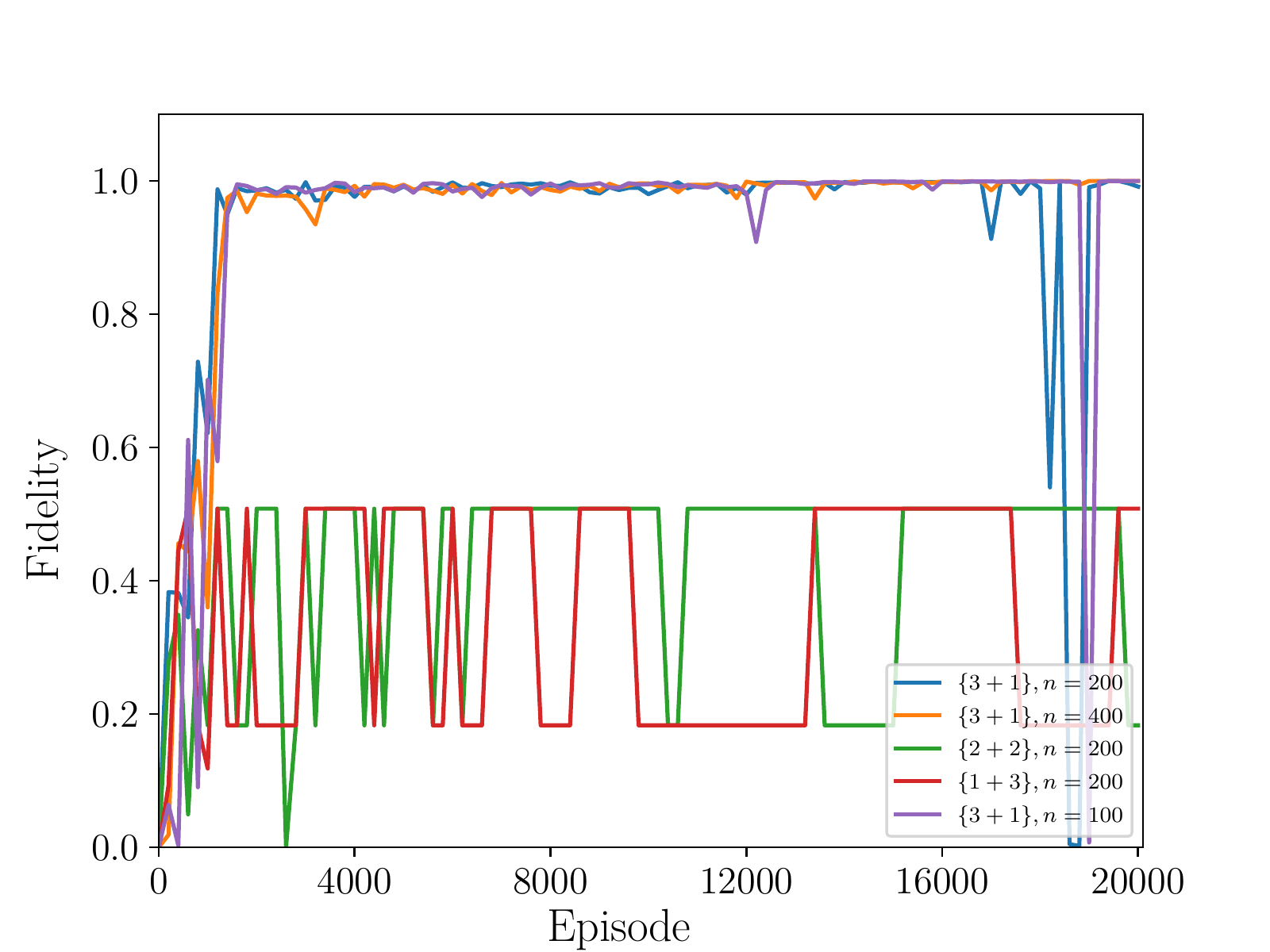}
 	\caption{\label{fig:arc}Learning curves of RL agent for Hadamard gate at $T=1$ with different neural network architectures. With different layer numbers $\{\textrm{encoder network + Advantage (Value) network} \}$ and neuron numbers $n$ of each architecture.}
 \end{figure}

\begin{figure}
 	\includegraphics[width=0.98\columnwidth{},keepaspectratio]{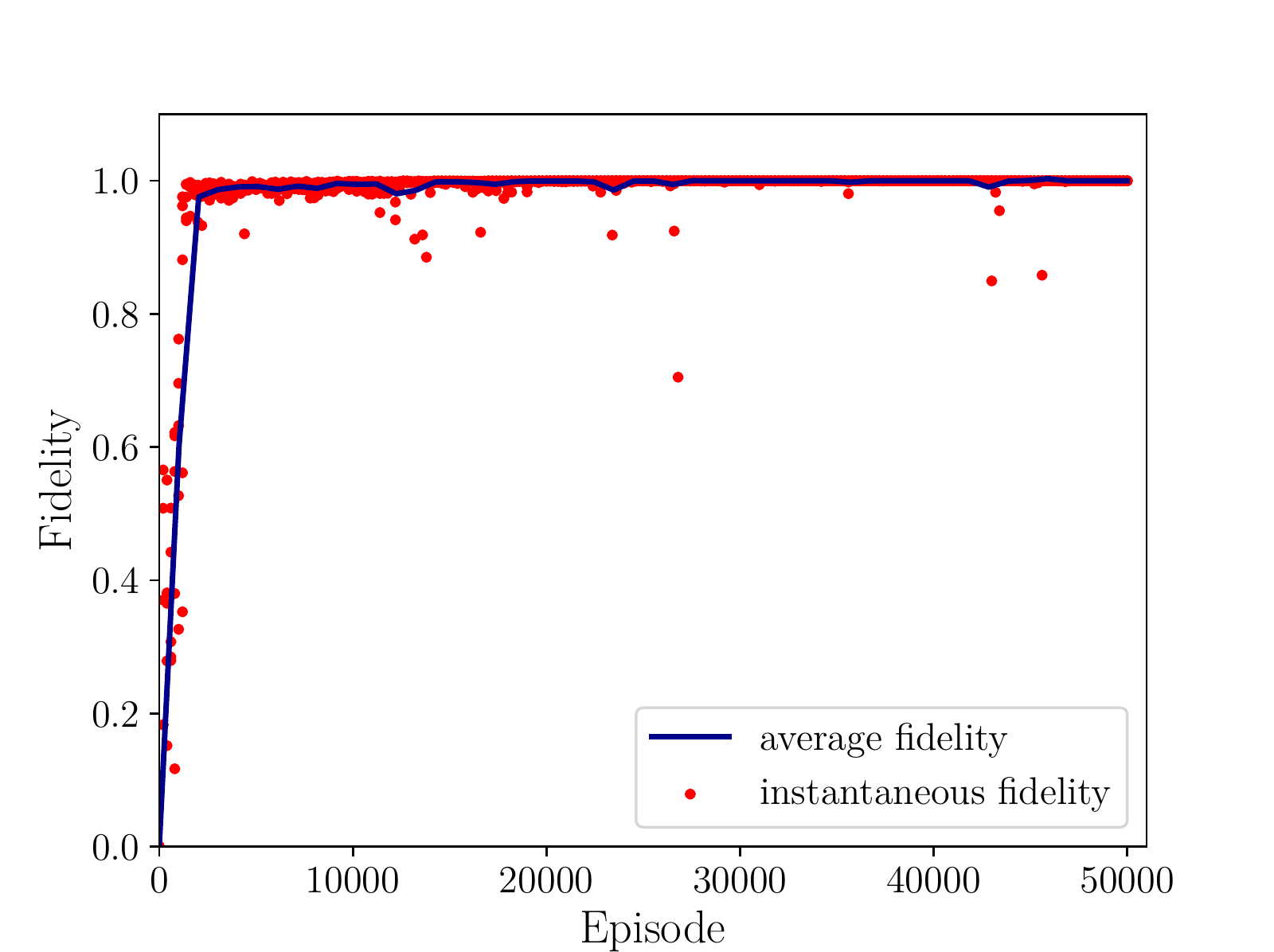}
 	\caption{\label{fig:trainh}Learning curves of the RL agent for
 		Hadamard gate at $T=1$. The red dots show the instantaneous
 		fidelity at every episode with 5 times, while the blue line the
 		average fidelity of the 5 agent.}
 \end{figure}

\begin{figure}
 	\includegraphics[width=0.98\columnwidth{},keepaspectratio]{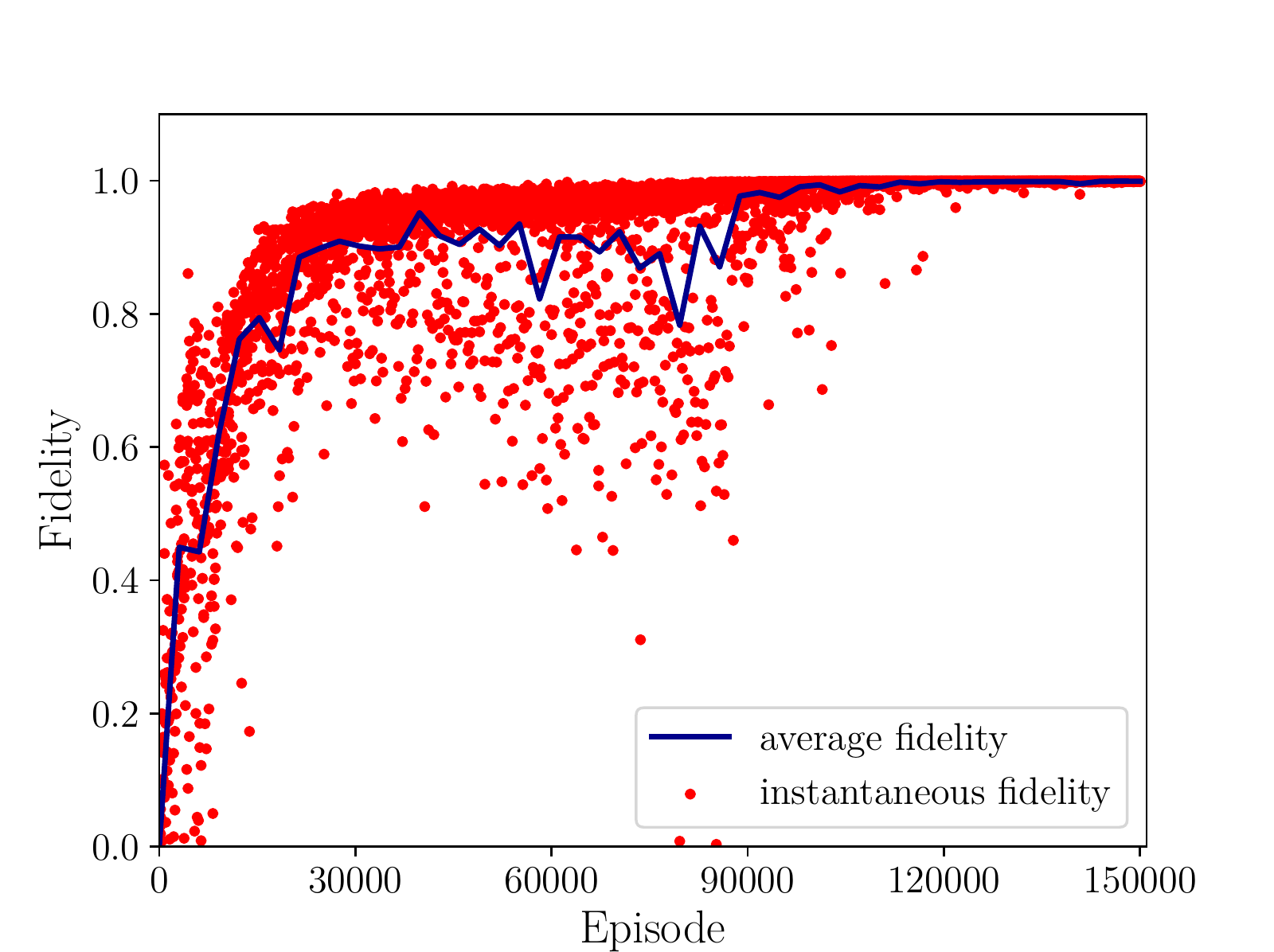}
 	\caption{\label{fig:trainc}Learning curves of the RL agent for CNOT
 		gate at $T=1.1$. The red dots show the instantaneous fidelity at
 		every episode with 5 times, while the blue line the average
 		fidelity of the 5 agent.}
 \end{figure}

 	Our RL agent makes use of a deep neural network to approximate
 the Q values for the possible actions of each state. The
 network (see Fig.~\ref{net}) consists of 4 layers of each
 sub-network. All layers have ReLU activation functions except
 the output layer which has linear activation. The
 hyper-parameters of the network are summarized in
 Table~\ref{para}. 
 	As shown in
 	Fig.~\ref{fig:arc}, the learning result highly depends on the
 	layer number of neural network. The computational
 	time is summarized in Table~\ref{time}. Notice
 	that the training time of two-qubit gate is from $6$ to $30$
 	times larger than that of one qubit gate. Among all
 	algorithms discussed in the paper, the resources needed by
 	our RL agent increase slowest. The learning
 curves for the two quantum gates are shown in
 Fig.~\ref{fig:trainh} and Fig.~\ref{fig:trainc}. All
 	algorithms are implemented with Python 3.6, and have been
 	run on two 14-core 2.60GHz CPU with 188 GB memory and four
 	GPUs.
 
\begin{table}
	\begin{threeparttable}
		\caption{Training Hyper-Parameters}
		
		\begin{tabular}{cc}
			\textrm{Hyper-parameter} &\textrm{Values} \\
			\hline
			Neurons in decoder network & $\{600,600,600\}$\\
			
			Neurons in advantage(value) network & $\{600,600,600,600\}$\\
			
			Minibatch size & \tnote{a} \\
			
			Replay memory size & 100000\\
			
			Learning rate & $0.001$\tnote{b}\\
			
			Update period & 100\\
			
			Reward decay $\gamma$ & 0.95\\
			
			Total episode & \tnote{c} \\
			
			\label{para}
		\end{tabular}
		\begin{tablenotes}\footnotesize
			\item [a] 72 for Hadamard gate problem, 128 for CNOT gate problem
			\item[b] With Adam algorithm
			\item[c] 50000 for Hadamard gate problem, 150000 for CNOT gate problem
		\end{tablenotes}
	\end{threeparttable}
\end{table}
 
\begin{table}
	\caption{Training time of different algorithms}
	
	\begin{tabular}{lll}
		\hline
		\textrm{Algorithm} &\textrm{Time} & \\
		\cline{2-3}
		&\textrm{Hadamard gate}&\textrm{CNOT gate}\\
		\hline
		GRAPE & $< 20s$ & about 7 min\\
		
		GA& about 20 min & about 10 h\\
		
		DE & about 40 min & about 18 h \\
		
		RL & about 5 h & about 31 h\\

		\label{time}
	\end{tabular}
	\begin{tablenotes}\footnotesize
		\item[a] The computation iterations is same with Fig~\ref{fig:compfid} and Fig~\ref{fig:compinf} 
	\end{tablenotes}
\end{table}
 
\FloatBarrier
\bibliographystyle{apsrev4-1}  \bibliography{gate_design.bib}
\end{document}